\documentclass[prd,twocolumn]{revtex4}
\usepackage{graphicx}
\newcommand{\be}{\begin{equation}}
\newcommand{\ee}{\end{equation}}
\newcommand{\bea}{\begin{eqnarray}}
\newcommand{\eea}{\end{eqnarray}}
\newcommand{\bwt}{\begin{widetext}}
\newcommand{\ewt}{\end{widetext}}
\begin{document}
\title{Colliders and Brane Vector Phenomenology}
\author{T.E. Clark}
\email[e-mail address:]{clark@physics.purdue.edu}
\affiliation{Department of Physics,\\
 Purdue University,\\
 West Lafayette, IN 47907-2036, U.S.A.}
\author{S.T. Love}
\email[e-mail address:]{love@physics.purdue.edu}
\affiliation{Department of Physics,\\
 Purdue University,\\
 West Lafayette, IN 47907-2036, U.S.A.}
\author{Muneto Nitta}
\email[e-mail address:]{nitta@phys-h.keio.ac.jp}
\affiliation{Department of Physics,\\
 Keio University,\\
 Hiyoshi, Yokohoma, Kanagawa, 223-8521, Japan}
\author{T. ter Veldhuis}
\email[e-mail address:]{terveldhuis@macalester.edu}
\affiliation{Department of Physics \& Astronomy,\\
 Macalester College,\\
 Saint Paul, MN 55105-1899, U.S.A.}
\author{C. Xiong
\footnote{Address after 1 September 2008: Department of Physics, University of Virginia, Charlottesville, VA. 22904-4714.  E-mail address: cx4d@virginia.edu.}}
\email[e-mail address:]{xiong@purdue.edu}
\affiliation{Department of Physics,\\
 Purdue University,\\
 West Lafayette, IN 47907-2036, U.S.A.}

\begin{abstract}
Brane world oscillations manifest themselves as massive vector gauge fields. Their coupling to the  Standard Model is deduced using the method of nonlinear realizations of the spontaneously broken higher dimensional space-time symmetries.  Brane vectors are stable and weakly interacting, and therefore escape particle detectors unnoticed.  LEP and Tevatron data on the production of a single photon in conjunction with missing energy are used to delineate experimentally excluded regions of brane vector parameter space.  The additional region of parameter space accessible to the LHC as well as a future lepton linear collider is also determined by means of this process.\newline
\end{abstract}
\maketitle

\section{Introduction}

At long wavelengths, the dynamics of the Nambu-Goldstone boson fields $\phi^i (x)$ associated with the spontaneous breakdown of translation invariance of the bulk due to an oscillating brane world \cite{Rubakov:1983bb} can be described by a nonlinear sigma model effective action \cite{Coleman:sm}.  This is the Nambu-Goto action \cite{Nambu:1974zg}, $\Gamma_{\rm N-G}$, of an invariant world volume made from the brane oscillation induced metric $g_{\mu\nu} =e_\mu^{~m}(\phi) \eta_{mn} e_\nu^{~n}(\phi)$ so that
\bea
\Gamma_{\rm N-G} &=& -\sigma \int d^4 x \det{e} \cr
 &=&-F_X^4 \int d^4 x \det{\sqrt{\delta^{ij} - \partial_\mu \phi^i \partial^\mu \phi^i/F_X^4}},
\label{N-GAction}
\eea
with the brane tension $\sigma= F_X^4$ and the co-volume having dimension $N$ so that $i=1,2,\ldots ,N$.  For dynamic space-time, the bulk gravitational field includes, besides the assumed world volume localized graviton, world volume gravi-vector fields, denoted $X_\mu^i$, related to the broken space translation directions.  Indeed, these vector fields combine with the Nambu-Goldstone bosons to become world volume (brane localized) massive Proca fields \cite{Clark:2006pd}-\cite{Kugo:1999mf}.  In the case of compact isotropic extra dimensions, for a flexible brane, one whose brane tension is less than the bulk gravitational scale, the coupling of the higher gravitational Kaluza-Klein modes to Standard Model fields is exponentially suppressed \cite{Kugo:1999mf}.  Hence the low energy effective description of the brane world dynamics includes the Standard Model fields, the graviton and these massive brane vector fields.  Indeed it has been shown in reference \cite{Geometry} using a Kaluza-Klein decomposition of the bulk gravitational fields that the gravi-vector fields become massive through the gravitational Higgs mechanism.  Further, the brane vector mass is enhanced or surpressed by the model dependent vacuum expectation value of the radion (dilaton) field, $M_X^2 = 2 e^{\frac{2N}{N+2}<\sigma>} \kappa^2 F_X^4$, where $\kappa$ is the world volume gravitational constant.  Hence the brane vector mass can range from small to large values conditional on the model dependent form of the bulk gravitational field.  Thus the brane vector mass is used as a phenomenological parameter in what follows, as are the brane tension and the various effective action coupling constants.

Since the brane vectors interact with Standard Model fields, they have the potential to be produced in collider experiments.  The purpose of this paper is to determine the constraints these experiments place on the parameters of the brane vector effective action \cite{Clark:2007wj, Colliders}.  Particular attention is focused on the production of a single photon plus 2 brane vectors.  For $N\geq 2$ isotropic codimensions or with an explicit $Z_2$ brane vector parity under which the vector is odd, as needed for $N=1$, the brane vectors must appear in pairs and thus are stable \cite{Clark:2007wj}. Consequently they appear as missing energy.  The single photon plus missing energy signature has been analyzed at LEP and the Tevatron. At present, the agreement with the Standard Model prediction for this process places an upper bound on the size of the contribution of new physics to the cross section.
This in turn places a bound on the brane vector parameter space.  Likewise the LHC and, in the future, a new lepton linear collider, here denoted by FLC (Future Linear Collider), will be able to access even larger portions of this parameter space as will be delineated in what follows.

In section 2 of this paper, the construction of the effective action is briefly reviewed \cite{Clark:2006pd}.  Since the embedded brane spontaneously breaks the bulk Poincar\'e symmetries the method of non-linear realizations has been used to determine the form of the effective action of the brane and its coupling to the Standard Model fields in a model independent manner.  The coefficients of the terms in the action are model dependent and are not determined by the coset method.  These parameters of the brane vector effective action are treated as phenomenologically determined couping constants and masses.  Specifically the interaction of the brane vectors with the Standard Model fields is expressed in terms of pairs of brane vectors coupled to the Standard Model energy-momentum tensor, and the brane extrinsic curvature and field strength coupled to the Standard Model weak hypercharge field strength as well as the Standard Model invariant Higgs mass term \cite{Clark:2007wj,Colliders}.  This last interaction leads to a direct coupling of the Higgs field to a pair of brane vector fields and hence to a new Higgs invisible decay mode whose rate can be comparable to any Standard Model decay rate of the Higgs \cite{higgsdecay}.  Being stable, the brane vectors are also dark matter candidates \cite{Clark:2007wj, darkmatter}.  Section 3 presents the cross section for production of a pair of brane vectors and a single photon from the annihilation of a positron and an electron in the case of LEP and the FLC.  The current bounds on the brane vector parameter space are exhibited in section 4 as obtained from LEP \cite{Clark:2007wj,Colliders}.  The regions of brane vector parameter space accessible to the FLC are presented there as well.  In section 5, the hadron collider cross section for a single photon and 2 brane vector production process is determined through the quark and anti-quark annihilation subprocess \cite{Colliders}.  The bounds on the brane vector parameter space obtained from the Tevatron-II data are presented in section 6 \cite{Colliders}.  In addition, the regions of parameter space that would be accessible to a continued running of the Tevatron up to an integrated luminosity of 6 ${\rm fb}^{-1}$ is described, as is the brane vector parameter space reach of the LHC \cite{Colliders}.  Concluding remarks are found in section 7.

\section{Brane Vector Effective Action}

Low energy theorems determine the form of the self coupling of Nambu-Goldstone bosons along with their coupling to other fields.  The effective action takes a non-linear sigma model Nambu-Goto form \cite{Clark:2007rn}.  For local symmetries, additional gauge fields are introduced to the model \cite{Clark:2006pd}.  In the unitary gauge, the Nambu-Goldstone bosons decouple per se and re-appear as the longitudinal modes of the now massive vector fields.  The formation of a brane in the bulk spontaneously breaks its translation and Lorentz transformation symmetries down to those of the world volume and its complement.  For simplicity consider the case of a domain wall in which the $D=5$ bulk Poincar\'e symmetries $ISO(1,4)$ are broken to the world volume Poincar\'e symmetries $ISO(1,3)$ \cite{Colliders}.  The symmetry generators can be denoted by their representation under the unbroken Lorentz group of the world volume, $SO(1,3)$.  Bulk translation symmetry generators $P^M = (P^\mu , Z=-P^{M=4})$ where $Z$ corresponds to the space translation symmetry generator broken by the domain wall.  The bulk angular momentum tensor can be decomposed as $M^{MN}=(M^{\mu\nu}, K^\mu =2 M^{4 \mu})$ with $K^\mu$ corresponding to the broken Lorentz transformations.  The covariant building blocks of the non-linear sigma model low energy effective action for this case are obtained from the coset elements $U(x)\in ISO(1,4)/SO(1,3)$
\be
U(x)= e^{ix^\mu  P_\mu} e^{i\phi(x)Z/F_X^2} e^{iv^\mu (x) K_\mu}.
\label{coset}
\ee  
The coset method applied to spontaneously broken space-time symmetries differs from the case of internal symmetries in that the coset includes the unbroken world volume space-time translations parameterized by the world volume coordinates $x^\mu$ along with the Nambu-Goldstone brane oscillation field $\phi (x)$ parameterizing broken space translations generated by $Z$ as would also be there in the internal symmetry case.  The other difference is that the coset contains the auxiliary Nambu-Goldstone boson fields $v^\mu (x)$ associated with the broken Lorentz transformation generators $K_\mu$ which are redundant.  A local broken Lorentz transformation parameterized by $v^\mu (x)$ is equivalent to a local broken space translation parameterized by $\phi (x)$.  Both sets of Nambu-Goldstone fields are needed in order to derive the correct field equations for the brane oscillations.  The auxiliary  Nambu-Goldstone boson fields $v^\mu (x)$ will appear, however, as Lagrange multiplier fields in the intermediate action and can by eliminated algebraically to obtain the Nambu-Goto action describing the brane oscillations into the covolume.  

The Maurer-Cartan one-form, $U^{-1}(x) \partial_\mu U(x)$, is the source of the covariant building blocks for the construction of the invariant, under all of the bulk symmetries, action
\bea
U^{-1}(x) \partial_\mu U(x) &=& i\left[e_\mu^{~m}(x) P_m +\nabla_\mu  \phi (x) Z/F_X^2  \right.  \cr
 & &\left. \qquad + \nabla_\mu v^{m} (x) K_m +\omega^{mn}_\mu (x) M_{mn}\right] .\cr
 & & 
\eea
The oscillations of the brane into the covolume are described by the Nambu-Goldstone field $\phi$.  It along with the auxiliary Nambu-Goldstone field $v^m$ induce a vierbein $e_\mu^{~m} (\phi,v^m)$ and a spin connection $\omega_\mu^{mn}=v^m \partial_\mu v^n -v^n \partial_\mu v^m +\cdots$ on the world volume.  The Maurer-Cartan one form coefficient of the broken bulk translation generator $Z$ yields the covariant derivative of the scalar Nambu-Goldstone field, $\nabla_\mu \phi (x)$.  The covariant derivative of $\phi$ can be used to express the auxiliary Nambu-Goldstone field $v_\mu$ in terms of $\phi$ by means of the covariant constraint 
$\nabla_\mu \phi =0$.  This allows the elimination of the auxiliary field $v_\mu = \partial_\mu \phi /F_X^2 + \cdots$ through the \lq\lq inverse Higgs mechanism" \cite{Ivanov:1975zq}.  Equivalently the Nambu-Goto action contains no derivatives of $v_\mu$ so that its Euler-Lagrange equation is purely algebraic and implies the same constraint.  Exploiting the constraint, the vierbein takes the form 
\be
e_\mu^{~m} = \delta_\mu^{~m} +\frac{\partial_\mu \phi \partial^m \phi}{\partial_\lambda \phi \partial^\lambda \phi}\left(\sqrt{1-\partial_\rho \phi \partial^\rho \phi /F_X^4} -1\right)
\ee
and the determinant of the vierbein simplifies to $\det{e} = \sqrt{1-\partial_\mu \phi \partial^\mu \phi /F_X^4}$ with the resulting Nambu-Goto action as in equation (\ref{N-GAction}).  The extrinsic curvature $K_{\mu\nu}$ of the brane is given by the covariant derivative of $v^\mu (x)$ which is the coefficient of the broken Lorentz transformation generator $K_m$.  After applying the constraint, its expansion begins with the second derivative of the brane oscillation coordinate: $K_{\mu\nu} = F_X e_\nu^{~m} \nabla_{\mu}v_m =F_X \partial_\mu v_\nu +\cdots =\partial_\mu \partial_\nu \phi /F_X +\cdots$.  $K_{\mu\nu}$ describes the rigidity or stiffness of the brane \cite{Polyakov:1986cs}.

The brane oscillation Nambu-Goto action \cite{Nambu:1974zg} along with the induced gravity action of the Standard Model fields \cite{Clark:2007rn} are invariant under the Poincar\'e symmetries of the bulk
\be
\Gamma_{\rm N-G} =  \int d^4 x \det{e} [-F_X^4 + {\cal L}_{\rm SM} (e)] ,
\ee
with ${\cal L}_{\rm SM} (e)$ the Standard Model Lagrangian coupled to the brane oscillations through the induced vierbein as in world volume general relativity.  For small brane oscillations relative to the tension, the action can be expanded in terms of derivatives of the brane oscillation fields.  Adding an explicit symmetry breaking mass $M_X$ for the scalar, the leading order interactions are obtained \cite{Creminelli:2000gh}, \cite{Cembranos:2004jp}
\be
\Gamma_{\phi\rm SM} = \int d^4 x \left[ \frac{1}{2 F_X^4}\partial^\mu \phi\partial^\nu \phi T_{\mu\nu}^{\rm SM}-\frac{M_X^2}{8F_X^4} \phi^2 \eta_{\mu\nu} T_{\mu\nu}^{\rm SM}\right],
\label{NGeffaction}
\ee
where $T_{\mu\nu}^{\rm SM}$ is the Standard Model symmetric energy-momentum tensor.  The ensuing phenomenology for massless \cite{Creminelli:2000gh} as well as massive \cite{Cembranos:2004jp} brane oscillation scalars has been extensively studied.

For the case that the gravitational fields are dynamic, the brane oscillation Nambu-Goldstone boson decouples and re-appears as the longitudinal degree of freedom of a gravi-vector field.  The brane oscillation is then described by this massive Proca field, denoted $X_\mu$, which is assumed to be localized on the world volume.  The coset method has been extended to include gravitational interactions \cite{Clark:2006pd}.  On the brane, the $ISO(1,4)$ symmetries of the bulk can be gauged by introducing the brane localized gravitational fields \cite{Clark:2006pd}
\bea
E_\mu (x)&=& E_\mu^{~m}(x) P_m +X_\mu (x) Z /F_X\cr
 & & \qquad + V_{~\mu}^m (x) K_{m} +\frac{1}{2}\gamma^{mn}_\mu (x) M_{mn} ,
\eea
with $E_\mu^{~m}(x)$ the dynamic gravitational vierbein and $\gamma^{mn}_\mu (x)$ the related spin connection on the brane.  In the domain wall with co-dimension one case, the single brane vector field $X_\mu (x)$ is the coefficient of the broken translation generator $Z$ while  $F_X$ is the effective brane tension.  $V_{~\mu}^m (x)$ is the field associated with the broken bulk Lorentz transformations.  In the effective action it will appear as an auxiliary field related to the second covariant derivative of the Nambu-Goldstone field $\phi$.  Since such derivatives already occur in the construction of the action it is redundant to introduce $V_{~\mu}^m (x)$.  The invariant action building blocks are the locally covariant Maurer-Cartan forms
\bea
U^{-1}(x)\left[ \partial_\mu + ie^{ix\cdot P} E_\mu (x) e^{-ix\cdot P}\right]U(x)\qquad\qquad\qquad\cr
  = i\left[e_\mu^{~m}P_m +\nabla_\mu \phi Z/F_X^2 + \nabla_\mu v^{m} K_m +\omega^{mn}_\mu M_{mn}\right] ,
\eea
where now the component one-forms depend on the dynamic gravitational fields as well as those induced by the Nambu-Goldstone fields.

The Nambu-Goldstone fields transform inhomogeneously under the broken space translations and broken Lorentz transformations.  The local transformation parameters can be used to transform to a partial unitary gauge in which $v_\mu (x)=0$ or a full unitary gauge in which $\phi (x)=0$ also.  In the partial unitary gauge, the Maurer-Cartan one-form coefficients become $e_\mu^{~m} (x) = \delta_\mu^{~m} +E_\mu^{~m}(x)+2\phi (x) V_{~\mu}^m (x)/F_X^2$ while the covariant derivative $\nabla_\mu \phi (x) = \partial_\mu \phi (x) + M_X X_\mu (x)$.  The spin connection has no induced term and is only given by the gravitational term, $\omega^{mn}_\mu (x) = \gamma^{mn}_\mu (x)$.  Similarly for the covariant derivative of $v^m$, $\nabla_\mu v^m (x)= V_{~\mu}^m (x)$.  According to the coset method, the action built from the Maurer-Cartan component forms is invariant under the larger symmetries of the bulk if it is invariant under the world volume unbroken local Poincar\'e symmetries.  The Standard Model fields can be similarly included in the bulk symmetry nonlinear realization construction according to their world volume unbroken Poincar\'e group representation.  Hence the invariant action involving the Standard Model fields will have the form of a world volume general coordinate invariant action \cite{Clark:2006pd}.  In the local case the covariant derivatives of the Nambu-Goldstone field $\phi$ are used to construct the brane vector invariant action.    Indeed they can be used to define the brane vector field strength tensor and the extrinsic curvature related tensor which can then be used as action building blocks.  The commutator of covariant derivatives yields the brane vector covariant field strength tensor $F_{\mu\nu} = [\nabla_\mu , \nabla_\nu ]\phi/M_X = \partial_\mu X_\nu -\partial_\nu X_\mu$.  The extrinsic curvature has leading terms given by the anti-commutator of covariant derivatives \cite{Geometry}.  As only these terms will be needed in the expansion of the effective action, the extrinsic curvature related symmetric 2 tensor is defined by this anti-commutator $K_{\mu\nu} \equiv 1/2 \{\nabla_\mu , \nabla_\nu \}\phi/M_X$.  Putting these elements together, the local bulk symmetry invariant action in the partial unitary gauge is given by the world volume general coordinate invariant action
\bwt
\bea
\Gamma &=& \int d^4 x \det{e}\left[ \Lambda + \frac{1}{2\kappa^2}R -\frac{1}{4}F_{\mu\nu} F^{\mu\nu}+ \frac{1}{2}\nabla^\mu \phi \nabla_\mu \phi +{\cal L}_{\rm SM} (e) \right. \cr
 & &\left. +\frac{\tau}{2F_X^4}
\nabla^\mu \phi \nabla^\nu \phi T^{\rm SM}_{\mu\nu}+\frac{M_X^2}{2F_X^4}\left(K_1 B_{\mu\nu} + K_2 \tilde{B}_{\mu\nu}\right) F^{\mu\rho} K_\rho^{~\nu} \right. \cr
 & &\left. + \frac{M_X^2}{2 F_X^4}\left[\lambda_1 K_{\mu\nu} K^{\mu\nu} +\lambda_2 F_{\mu\nu} F^{\mu\nu} +\lambda_3 F_{\mu\nu} \tilde{F}^{\mu\nu} \right] (\Phi^\dagger \Phi -\frac{v^2}{2}) \right] ,
\eea
\ewt
where $M_X$ is the brane vector mass and $\tau$, $K_1$, $K_2$, $\lambda_1$, $\lambda_2$ and $\lambda_3$ are coupling constants and $v$ is the Higgs vacuum expectation value.  The mixed extrinsic curvature and field strength tensor coupling terms involve the $U(1)$ hypercharge field strength tensor $B_{\mu\nu}$ and its dual $\tilde{B}_{\mu\nu}$.  The terms coupling to the invariant Higgs weak doublet $\Phi$ composite, $\Phi^\dagger \Phi$, can lead to invisible Higgs decay as discussed in reference \cite{higgsdecay}.    

The remaining broken local space translation symmetry can be used to transform to the full unitary gauge in which $\phi (x)=0$ also.  In this gauge only the physical degrees of freedom remain.  The purely gravitational world volume vierbein is $e_\mu^{~m}(x) = \delta_\mu^{~m} +E_\mu^{~m}(x)$ and its effects will be ignored in what follows.  The covariant derivative of the Nambu-Goldstone scalar field is the canonical dimension brane vector field $X_\mu$ times the brane vector mass $M_X$, $\nabla_\mu \phi (x)= M_X X_\mu (x)$.  The extrinsic curvature related tensor becomes $K_{\mu\nu}= (1/2)(\partial_\mu X_\nu +\partial_\nu X_\mu)+\cdots$.  Extending this derivation to co-dimension $N$ in which case there are $N$-species of brane vector, $X_\mu^i$, and ignoring the purely world volume gravitational interactions, the effective action detailing the interaction of the brane vectors with the Standard Model fields in leading order in $1/F_X$ is obtained \cite{Clark:2007wj, Colliders}
\bwt
\bea
\Gamma_{\rm Effective}&=&\int d^4 x \left[{\cal L}_{\rm SM}(\eta) -\frac{1}{4}F_{\mu\nu}^i F_i^{\mu\nu} +\frac{1}{2}M_X^2 X_\mu^i X_i^\mu \right.\cr 
 & & \left. +\frac{\tau}{2}\frac{M_X^2}{F_X^4}X^\mu_i T^{\rm SM}_{\mu\nu} X^\nu_i +\frac{M_X^2}{2 F_X^4}\left(K_1 B_{\mu\nu} +K_2 \tilde{B}_{\mu\nu} \right) F^{i\mu\rho} K_\rho^{i\nu}\right. \cr
 & &\left. +\frac{M_X^2}{2 F_X^4}\left[  \lambda_1 K_{\mu\nu}^i K^{i\mu\nu}+\lambda_2 F_{\mu\nu}^i F^{i\mu\nu} +\lambda_3 F_{\mu\nu}^i \tilde{F}^{i\mu\nu}\right] (\Phi^\dagger \Phi -\frac{v^2}{2})\right] ,
\label{effaction}
\eea
\ewt
where only the leading form of the extrinsic curvature related tensor is needed $K^i_{\mu\nu}= 1/2(\partial_\mu X^i_\nu +\partial_\nu X^i_\mu)$ and ${\cal L}_{\rm SM}(\eta)$ is the flat space Standard Model lagrangian.  

The covolume $SO(N)$ symmetry is taken to be in the Higgs phase, hence, the associated gauge fields are massive and not considered here \cite{Clark:2006pd}.  Once the $SO(N)$ symmetry is broken the component fields of the $N$-plet brane vector will have different masses and effective brane tension scales.  The brane vectors are taken to have the same mass $M_X$ and effective brane tension $F_X$ for the above initial phenomenology.  Similarly, the bilinear $X$ coupling, $\nabla^\mu \phi^i \nabla^\nu \phi^i \rightarrow (M_X^2 ) X^{i\mu} X^{i\nu}$, is to any $SU(3)\times SU(2) \times U(1)$ invariant with its own characteristic dimensionless coupling constant.  The scale for all terms is set by the effective brane tension $F_X$.  These coupling constants also have been chosen to be equal.  As such the brane vectors couple pairwise to the Standard Model energy-momentum tensor with an overall coupling strength $\tau$.  This has the practical advantage of comparison to the global symmetry branon case where the Nambu-Goldstone brane oscillation scalar $\phi^i$ must couple pairwise derivatively to $T_{\mu\nu}^{\rm SM}$.  The derivative brane vector coupling, however, arises from a combination of terms made from the extrinsic curvature related tensor and the field strength tensor.  The terms involving the field strength tensor provide coupling to the Standard Model invariants that vanish in the global branon limit.  The three parameters $\lambda_1$, $\lambda_2$ and $\lambda_3$ describe the direct coupling to the Higgs particle.  The Standard Model masses are neglected in what follows so that the coupling to the trace of the energy-momentum tensor, $(1/F_X^4 )\nabla^\rho \phi^i \nabla_\rho \phi^i \eta^{\mu\nu} T^{\rm SM}_{\mu\nu}\rightarrow (M_X^2 /F_X^4) X^{i\rho} X^i_\rho \eta^{\mu\nu} T^{\rm SM}_{\mu\nu}$, is ignored.  In addition the original isotropic covolume invariance ($SO(N>1)$), or a $Z_2$ brane vector parity invariance, ($X_\mu^i \rightarrow -X_\mu^i$), needed in the $N=1$ case, requires the brane vector to occur in pairs.  Hence, terms linear (as well as all odd powers) in the brane vector field are absent from the effective action.  The brane vector particles are stable.

\section{Single Photon and $XX$ Production:\newline
LEP and the FLC}

The effective action coupling the brane vectors to the Standard Model fields determines their possible production in collider experiments.  The production of a single photon and two brane vectors, which escape the detector and appear as missing energy, is the focus of this paper.  The single photon plus missing energy experiments at LEP and the Tevatron are consistent with the expected number of events from the Standard Model.  This implies that the cross section coming from new physics must be constrained so as not to contribute measureably above the Standard Model background.  The constraints on the brane vector parameter space obtained from LEP were initially reported in reference \cite{Clark:2007wj} and will be elaborated upon here.  As well the regions of parameter space accessible to a future lepton collider, FLC, can also be estimated by means of the single photon plus missing energy experiments.  From the effective action (\ref{effaction}), the Feynman diagrams contributing to the production process in leading order are shown in Fig. 1 with the Standard Model fermion $f$ the electron and anti-fermion $\bar{f}$ the positron.  

The differential cross section for spin averaged incoming electon and positron annihilation producing a photon and 2 brane vectors, summed over all polarizations and $X$ species $i=1,2,\ldots, N$ is found to be \cite{Colliders}
\bwt
\bea
\frac{d\sigma_{\gamma XX}}{dx d\cos{\theta}}&=&\frac{\alpha}{4\pi}\frac{1}{15,360 \pi}
\left[\frac{N}{F_X^8}\right] \frac{s\sqrt{s(1-x) - 4M_X^2}}{\sqrt{s(1-x)}}\frac{1}{x\sin^2{\theta}}\left(4-4x+2x^2-x^2\sin^2{\theta}\right)\cr
 &\times& \left\{ \tau^2 \left( 1-x+x^2\sin^2{\theta}\right) \left(\left[s(1-x) - 4M_X^2\right]^2 +20 M_X^2 \left(s(1-x) +2M_X^2\right) \right) \right. \cr
 & &\left. +\left[K_1^2 s(1-x) +K_2^2 \left(s(1-x) -4M_X^2\right) \right] (SM) \right. \cr
 & &\left. \qquad\qquad\qquad\qquad\times\left[80M_X^2  s(1-x)^2 \left(s(1-x)-4M_X^2\right) \right]\right\} ,
\label{xsection1}
\eea
\ewt
where the electromagnetic fine structure constant $\alpha$ is evaluated at the $W$ mass.  The Standard Model factor, $(SM)$, arises from the photon or $Z$ exchange in the last 2 Feynman graphs in Fig. 1 and takes the form
\bea
(SM) &=& \pi \alpha \left\{ \frac{\cos^2{\theta_W}}{(k^2)^2}+ \frac{1}{(k^2 -M_Z^2)^2 +M_Z^2 \Gamma_Z^2} \right.\cr
 &\times&\left. \left[\frac{1}{\cos^2{\theta_W}} \left(\frac{1}{16} + \left(\sin^2{\theta_W} -\frac{1}{4}\right)^2 \right)\right. \right. \cr
 & &\left.\left. +2\cos^2{\theta_W} \left(\sin^2{\theta_W} -\frac{1}{4}\right)\frac{(k^2 -M_Z^2)}{k^2}\right]\right\},\cr
 & & 
\label{SM}
\eea
with $\theta_W$ the Weinberg angle and $M_Z$ and $\Gamma_Z$ the mass and width of the $Z$ gauge boson, respectively.  Note that there is no interference between any of the different interaction terms, that is no mixed factors of the energy-momentum tensor coupling terms $\tau$ and the extrinsic curvature related interactions $K_1$ and $K_2$.

\begin{figure}
\begin{center}
\includegraphics[height=3.0in]{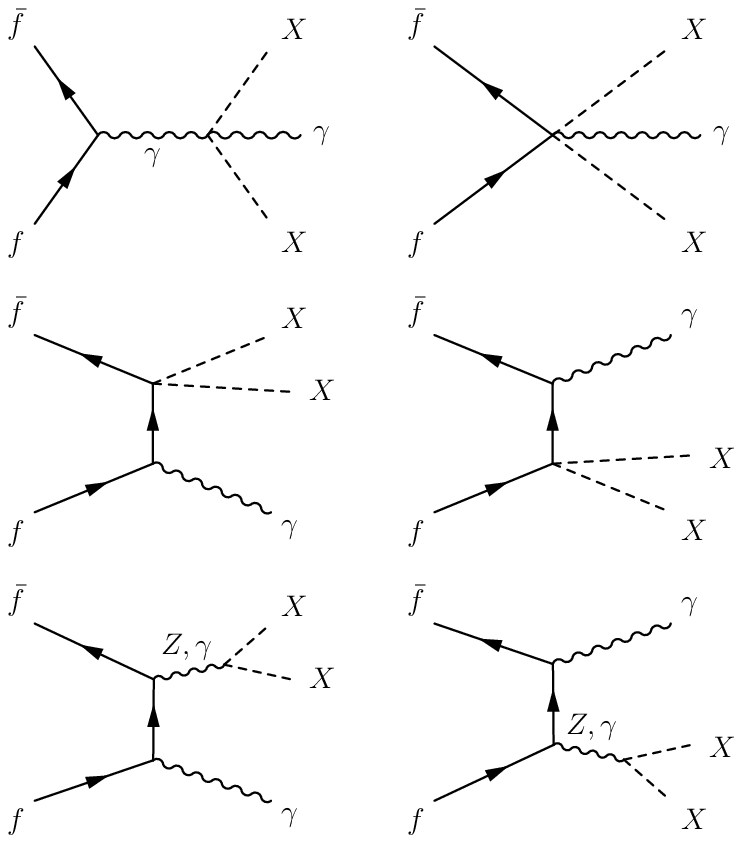}
\end{center}
\caption{The Feynman graphs contributing to the single photon and missing energy brane vector production.  The top four graphs involve $T^{\rm SM}_{\mu\nu}$ while the bottom two graphs depend on the coupling to $K^i_{\mu\nu}$.}
\end{figure}

The differential cross section is given in terms of the center of mass energy $\sqrt{s}$ with the positron and electron each having the beam energy $E_{\rm Beam}= \sqrt{s}/2$.  The single photon carries away the fraction $x$ of beam energy, $E_\gamma = x E_{\rm Beam}$, and emerges at a polar angle $\theta$ with respect to the positron beam axis.  The photon transverse energy is $E_{\gamma \rm T}= E_\gamma \sin{\theta}$.  The sum of the $X$ particles' 4-momenta squared is $k^2=(k_1+k_2)^2=s(1-x)$.  The total cross section is then obtained by directly integrating over the photon energy fraction $x$ and its polar angle $\theta$ as \cite{Clark:2007wj, Colliders}
\be
\sigma_{\gamma XX} = \int_{\cos{\theta_{\rm min}}}^{\cos{\theta_{\rm max}}} d(\cos{\theta})\int_{x_{\rm min}}^{x_{\rm max}} dx \frac{d\sigma_{\gamma XX}}{dx d\cos{\theta}} .
\label{totalxsection1}
\ee
Following the LEP-II experimental analysis of the single photon plus missing energy signature, the photon polar angle is integrated over the range $|\cos{\theta}|\leq 0.97$ while the photon transverse energy has minimum and maximum cuts given by $0.04  \leq E_{\gamma \rm{T}}/E_{\rm Beam}\leq 0.60$ \cite{LEP}.  Kinematically in order to produce 2 brane vectors the sum of their 4-momenta squared must be greater than $4M_X^2$.  Hence the photon energy fraction $x\leq 1-4M_X^2/s$ and the bounds on the $x$ integration, $x_{\rm min\atop{max}}$, are the lesser of $1-4M_X^2/s$ or ${0.04 \atop{0.60}}/\sin{\theta}$.  The same cuts will be employed in the analysis of the FLC reach.

\section{Brane Vector Parameter Space:\newline
Lepton Colliders}

The data from LEP is in agreement with the Standard Model for the annihilation process $e^+ ~e^- \rightarrow \gamma + \rlap{E}{/}~$ \cite{LEP}.  For there to be a discovery of new physics contributing to this missing energy process the number of such events is taken to be 5 sigma above the Standard Model background.  Hence the cross section for production of 2 brane vectors and a single photon, $e^+ e^- \rightarrow \gamma + XX\rightarrow \gamma + \rlap{E}{/}$, must be constrained to be less than this 5-$\sigma$ discovery cross section,  $\sigma_{\rm Discovery} $, so that
\bea
\sigma_{\gamma XX} {\cal L}({\rm LEP}) &\leq & 5 \sqrt{\sigma_{\rm SM~Bkgrnd} {\cal L}{\rm(LEP)}} \cr
 &\equiv & \sigma_{\rm Discovery} {\cal L}({\rm LEP}) ,
\eea
where ${\cal L}({\rm LEP})$ is the integrated luminosity for the LEP-II experimental run.  This experimental constraint then translates into an excluded region of parameter space.  The allowed values of the brane tension are bounded from below by a function of the brane vector mass and the coupling constants' values \cite{Clark:2007wj, Colliders}.
\vspace*{0.1in}
\begin{center}
\begin{tabular}{|cccc|}
\hline
Collider&$\sqrt{s}$ (GeV)&${\cal L}~{\rm (pb^{-1})}$&$\sigma_{\rm Discovery}$ (pb)\\\hline\hline
LEP-II&206&138.8&0.45\\
FLC&500&$2\times 10^5$&0.012\\
\hline
\end{tabular}
\end{center}
\begin{center}
{\bf Table 1.}  The colliders, their center of mass energies,\\ integrated
luminosities and discovery cross-sections.
\end{center}
As can be seen from the differenctial cross section the dependence of this line of exclusion for $F_X$ on the number of extra dimensions is very mild, varying as $N^{1/8}$.  Hence only co-dimension $N=1$ values will be plotted.  The bounds on $F_X$ and $M_X$ are shown in Fig. 2 for different values of the coupling to the energy-momentum tensor, given by the coupling constant $\tau$, and to the extrinsic curvature related terms with coupling constants $K_1$ and $K_2$.  Fig. 3 shows excluded and allowed regions of $K_1$ and $K_2$ parameter space for fixed values of the brane tension $F_X$ and brane vector mass $M_X$ for $\tau =1$.  Table 1 summarizes the center of mass energies, the integrated luminosities and backgrounds for each collider \cite{LEP}.
\begin{figure}
\begin{center}
\includegraphics[height=2.2in]{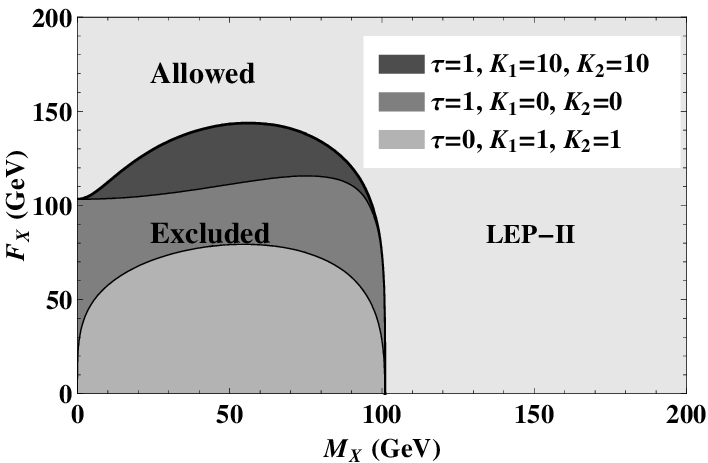}
\end{center}
\caption{LEP-II excluded regions of brane vector parameter space for fixed values of $\tau$, $K_1$ and $K_2$.}
\end{figure}
\begin{figure}
\begin{center}
\includegraphics[height=2.0in]{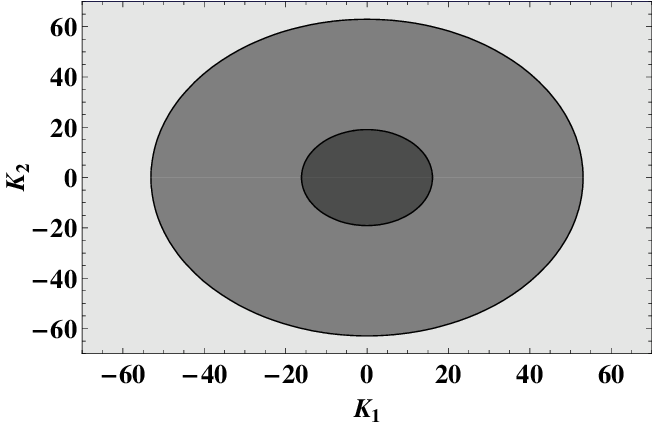}
\end{center}
\caption{LEP-II excluded regions of brane vector parameter space for fixed values of $F_X$, $M_X$ and $\tau =1$.  The outer lightest grey areas are excluded regions of parameter space while the 2 inner grey regions are allowed.  The innermost area corresponds to $F_X =150$ GeV and $M_X=50$ GeV.  The medium grey middle region corresponds to $F_X =200$ GeV and $M_X=50$ GeV.}
\end{figure}

For large values of the energy compared to the brane vector mass, the longitudinal mode will dominate the scattering.  Using the equivalence theorem \cite{equivthm}, the longitudinal brane vector cross section can be accurately described by that of the Nambu-Goldstone boson \cite{Kugo:1999mf}, \cite{Cembranos:2004jp}, \cite{Creminelli:2000gh}.  The Nambu-Goldstone boson differential cross section as calculated from the top 4 Feynman diagrams in Fig. 1 with Feynman rules obtained from the effective action Eq.(\ref{NGeffaction}) is given by
\bea
\frac{d\sigma^{\rm N-G}_{\gamma XX}}{dx d\cos{\theta}}&=&\frac{\alpha}{4\pi}\frac{1}{15,360 \pi}
\left[\frac{N}{F_X^8}\right] \frac{s\sqrt{s(1-x) - 4M_X^2}}{\sqrt{s(1-x)}}\cr
 &\times& \frac{1}{x\sin^2{\theta}}\left(4-4x+2x^2-x^2\sin^2{\theta}\right)\cr
 &\times& \left( 1-x+x^2\sin^2{\theta}\right) \left[s(1-x) - 4M_X^2\right]^2 .\cr
 & & 
\label{xsectionNGBoson}
\eea
The total cross section is obtained by integrating using the same cuts as in Eq. (\ref{totalxsection1}).  This effect can be seen in Fig. 4 for $\tau =1$ and $K_1 =0=K_2$ for small mass values.  The Nambu-Goldstone boson (branon) degree of freedom excluded/allowed region is shown along with the full brane vector constraints.  For small values of $M_X$ the lines of exclusion approach each other.  For larger values of the brane vector mass, the full brane vector cross section is required.  
\begin{figure}
\begin{center}
\includegraphics[height=2.2in]{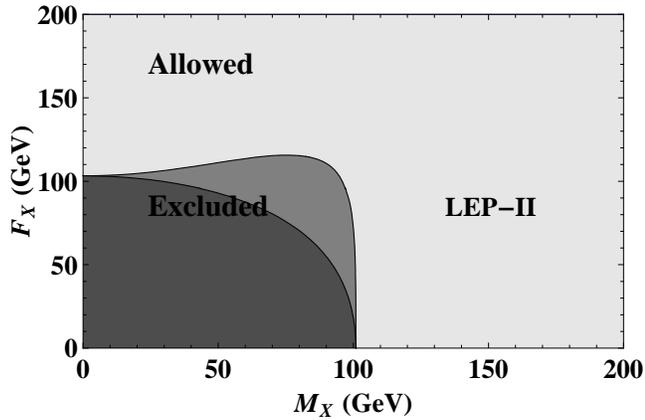}
\end{center}
\caption{Comparison of the branon and brane vector exclusion regions.  For small mass $M_X$ the equivalence theorem applies.  Branon excluded region in darker grey shading superimposed over the brane vector excluded region.}
\end{figure}

Larger regions of brane vector parameter space will be accessible to the FLC.  A conservative estimate of the discovery cross section sensitivity of the FLC is found by scaling the LEP-II background by the FLC luminosities.  Requiring a new physics discovery to be 5-$\sigma$ above this background yields
\be
\sigma_{\rm Discovery} ({\rm FLC}) = \sigma_{\rm Discovery} ({\rm LEP}) \sqrt{\frac{{\cal L}({\rm LEP})}{{\cal L}({\rm FLC})}} = 0.012 ~{\rm pb}.
\ee
\begin{figure}
\begin{center}
\includegraphics[height=2.2in]{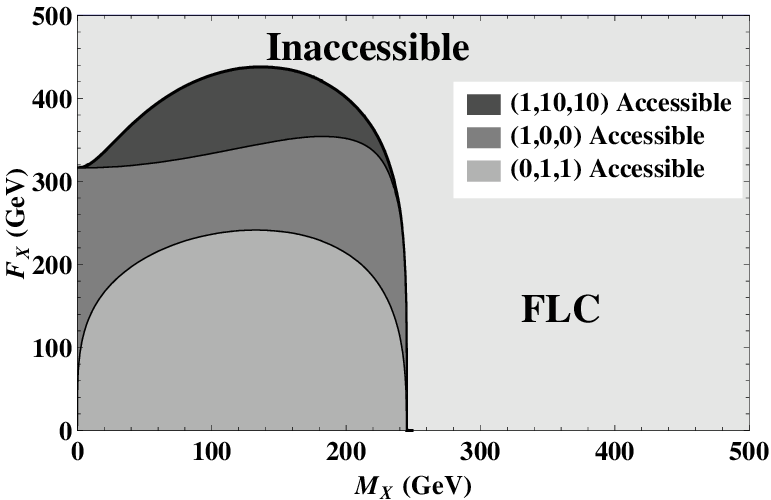}
\end{center}
\caption{FLC accessible regions of brane vector parameter space for fixed values of $\tau$, $K_1$ and $K_2$.  The triplets $(\tau, K_1 , K_2)$ denote the values of the corresponding coupling constants.}
\end{figure}
\begin{figure}
\begin{center}
\includegraphics[height=2.0in]{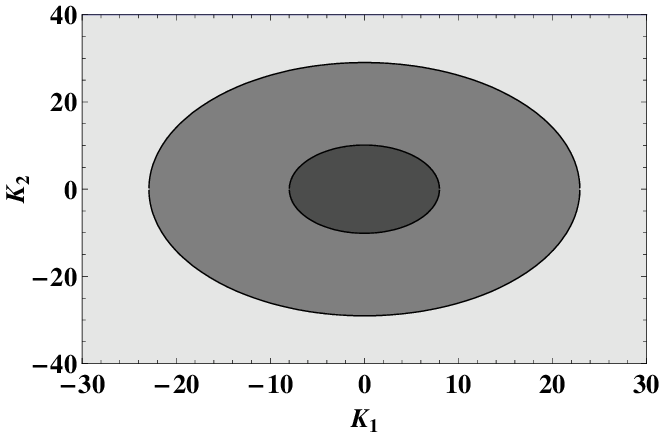}
\end{center}
\caption{FLC accessible regions of brane vector parameter space for fixed values of $F_X$, $M_X$ and $\tau =1$.  The outer lightest grey areas are accessible regions of parameter space while the 2 inner grey regions are inaccessible.  The innermost area corresponds to $F_X =400$ GeV and $M_X=140$ GeV.  The medium grey middle region corresponds to $F_X =500$ GeV and $M_X=140$ GeV.}
\end{figure}
Hence the region of accessibility of the brane vector parameter space probed by the FLC using the $e^+ e^- \rightarrow \gamma + XX~$ cross section is 
\be
\sigma_{\gamma XX} \geq \sigma_{\rm Discovery} ({\rm FLC}),
\ee
and conversely, if $\sigma_{\gamma XX} \leq \sigma_{\rm Discovery} ({\rm FLC})$ the corresponding region of parameter space will be inaccessible to the FLC.  Various slices of parameter space displaying the regions of accessibility and inaccessibility are shown in Fig. 5 for fixed values of $\tau$, $K_1$, $K_2$ and $N=1$ and in Fig. 6 for fixed values of $\tau=1$, $F_X$, $M_X$ and $N=1$.\newline

\section{Single Photon and $XX$ Production:\newline
Tevatron and the LHC}

Tevatron data from the single photon plus missing energy experiments further delineates the allowed regions of parameter space, as will results from the LHC.  For all these hadronic experiments, the subprocess of relevance is the annihilation of quark and anti-quark to produce a single photon and a pair of brane vector particles.  From the effective action, equation (\ref{effaction}), the Feynman diagrams contributing to the production process to leading order are again shown in Fig. 1 this time with $f$ denoting a quark and $\bar{f}$ an anti-quark.

The differential cross section for spin averaged incoming quark and anti-quark annihilation producing a photon and 2 brane vectors, summed over all polarizations and $X$ species $i=1,2,\ldots, N$ is found to be \cite{Colliders}
\bwt
\bea
\frac{d^2 \sigma_{\gamma XX}}{dk^2 dt}&=& \frac{\alpha}{4\pi}\frac{1}{15,360 \pi}
\left[\frac{N}{F_X^8}\right] \frac{1}{\hat{s}^3 u t}\frac{\sqrt{k^2 - 4M_X^2}}{\sqrt{k^2}} \left[2\hat{s} k^2 +u^2 +t^2\right] \cr
 &\times& \left\{ \tau^2 \left( \hat{s} k^2 +4 u t\right) \left(\left[k^2 - 4M_X^2\right]^2 +20 M_X^2 \left(k^2 +2M_X^2\right) \right) \right. \cr
 & &\left. +\left[K_1^2 k^2 +K_2^2 \left(k^2 -4M_X^2\right) \right] (SM) \left[80 \hat{s} \left(k^2\right)^2 (k^2-4M_X^2)M_X^2 \right]\right\} ,\cr
 & & 
\label{xsection2}
\eea
\ewt
where the Standard Model factor, $(SM)$, is given by equation (\ref{SM}), but now with $k=k_1+k_2$, where $k_1$ and $k_2$ are the brane vector 4-momenta.  As previously noted, there is no interference between the energy-momentum tensor coupling terms and the extrinsic curvature related interactions.

The cross section, Eq. (\ref{xsection2}), is obtained by expressing Eq. (\ref{xsection1}) in terms of Mandelstam variables $\hat{s} = (p_1 +p_2)^2$, $t=(p_1-q)^2$ and $u=(p_2-q)^2$, so that $\hat{s} + t+ u = k^2 +q^2 +p_1^2 +p_2^2 =k^2$, where the second equality follows from neglecting the quark masses.  These are given in terms of the momenta of the quark, $p_2$, the anti-quark, $p_1$, the photon, $q$, and the 2 brane vectors, $k_1$ and 
$k_2$.  In the case of the Tevatron collider, the proton carries the beam energy, $E_{\rm Beam} =\sqrt{s}/2$, which is half the center of mass energy $\sqrt{s}$, while the subprocess (anti-)quark carries a fraction $x$ of the proton energy, $x\sqrt{s}/2$.  The anti-proton also has half the center of mass energy while the (anti-)quark carries a fraction $y$ of the anti-proton energy, $y\sqrt{s}/2$.  Likewise in the case of the LHC, each proton has half the center of mass energy while one (anti-)quark carries fraction $x$ of that energy while the other (anti-)quark carries fraction $y$ of its proton's energy.  Neglecting the quark masses, it follows that for the hadron colliders $\hat{s}=2 p_1^\mu p_{2\mu}$ is related to the center of mass energy by $\hat{s}= x y s$.

The quark and anti-quark carry only fractions of the beam energy.  Hence, to find the total cross section, the differential cross section must be multiplied by the quark distribution functions and integrated over the range of energies of the quarks, that is the $x$ and $y$ energy fractions, for their kinematically allowed regions.   The photon's polar angle $\theta$ from the beam axis is given in terms of the pseudorapidity $\eta = -\ln\tan{(\theta/2)}$.  The cuts on the polar angle of the photon determine a minimum and maximum pseudorapidity.  This then yields minimum and maximum $t$ integration limits: $t_{{\rm min}\atop {\rm max}}=(k^2 -\hat{s})x[1-\tanh{\eta_{{\rm min}\atop {\rm max}}}]/[(y+x)+(y-x)\tanh{\eta_{{\rm min}\atop {\rm max}}}]$.  The transverse energy of the photon, $E_{\gamma \rm T} = E_\gamma \sin{\theta}$, has a minimum cut, $E_{\gamma \rm T}^{\rm min}$, which in turn yields a maximum for the $k^2$ variable, $k^2_{\rm max} = \hat{s}(1 -2E_{\gamma \rm T}^{\rm min}/\sqrt{\hat{s}})$.  The minimum integration limit to produce 2 $X$ particles is given by $k^2_{\rm min} = 4 M_X^2$.

The total cross section is obtained by integrating equation (\ref{xsection2}) over the appropriate $t$ and $k^2$ ranges and with the parton distribution functions, denoted by $f(x, y; \hat{s})$, over the fractional (anti-)quark energies given by $x$ and $y$ with $\hat{s}= x y s$.  The energy fractions $x$ and $y$ have the lower integration limits given by $x_{\rm min} = \hat{s}_{\rm min} /s$ and $y_{\rm min} = x_{\rm min}/x$ with $\hat{s}_{\rm min} = 2 E_{\gamma \rm T}^{{\rm min}2} +4 M_X^2 + 2 E_{\gamma \rm T}^{\rm min} \sqrt{E_{\gamma \rm T}^{{\rm min}2} +4 M_X^2}$.  The CTEQ-6.5M quark distribution functions \cite{CTEQ} appropriate for the energy range and process are used to obtain the form of the total cross section for the hadronic colliders \cite{Colliders}
\bea
\sigma_{\gamma XX}^{\rm Hadron}&=& \int_{x_{\rm min}}^1 dx \int_{y_{\rm min}}^1 dy f(x, y; \hat{s})\cr
 & & \qquad\times  \int_{k^2_{\rm min}}^{k^2_{\rm max}} dk^2 \int_{t_{\rm min}}^{t_{\rm max}} dt \frac{d^2 \sigma_{\gamma XX}}{dk^2 dt} .
\eea
The quark distribution function employed for the Tevatron $p\bar{p}$ collisions is
\bea
f_{\rm TeV}(x, y; \hat{s})=\qquad\qquad\qquad\qquad\qquad\qquad\qquad \cr
\frac{1}{3}\left(\frac{2}{3}\right)^2 \left[ u_p (x,\hat{s}) \bar{u}_{\bar{p}} (y, \hat{s}) + \bar{u}_p (x,\hat{s}) {u}_{\bar{p}} (y, \hat{s})\right] \qquad\cr
 + \frac{1}{3}\left(-\frac{1}{3}\right)^2 \left[ d_p (x,\hat{s}) \bar{d}_{\bar{p}} (y, \hat{s}) + \bar{d}_p (x,\hat{s}) {d}_{\bar{p}} (y, \hat{s})\right] ,
\eea
while the quark distribution function used for the LHC $pp$ collisions is
\bea
f_{\rm LHC}(x, y; \hat{s}) =\qquad\qquad\qquad\qquad\qquad\qquad\qquad \cr 
\frac{1}{3}\left(\frac{2}{3}\right)^2 \left[ u_p (x,\hat{s}) \bar{u}_{{p}} (y, \hat{s}) + \bar{u}_p (x,\hat{s}) {u}_{{p}} (y, \hat{s})\right] \qquad\cr
 + \frac{1}{3}\left(-\frac{1}{3}\right)^2 \left[ d_p (x,\hat{s}) \bar{d}_{{p}} (y, \hat{s}) + \bar{d}_p (x,\hat{s}) {d}_{{p}} (y, \hat{s})\right] .
\eea
Here $u_p$ denotes the fraction of up quark in the proton, $\bar{u}_p$ denotes the fraction of anti-up quark in the proton and so on.  For annihilation, the overall 1/3 is the probability the quarks have the same color and the distribution functions $f$ include the electric charge coupling of the quarks in units of $e$.  The pseudorapidity has the range $|\eta|\leq 1.0$ for both experiments and the minimum transverse energy of the photon is $E_{\gamma \rm T}^{\rm min} = 45$ GeV for the Tevatron \cite{Tevatron} and is scaled to $E_{\gamma \rm T}^{\rm min} = 350$ GeV for the LHC.

\section{Brane Vector Parameter Space:\newline
Hadron Colliders}

The data from the Tevatron is also in agreement with the Standard Model for the annihilation process $p\bar{p} \rightarrow \gamma + \rlap{E}{/}~$.  Thus the analysis of the constraints imposed on the brane vector parameter space will be analogous to that used in the LEP case \cite{Clark:2007wj, Colliders}.  For a discovery of new physics contributing to this missing energy process we take that the number of such events will correspond to a 5 sigma signal above the Standard Model background.  Hence the cross section for production of 2 brane vectors and a single photon, $p\bar{p} \rightarrow \gamma + XX\rightarrow \gamma + \rlap{E}{/}~$, must be constrained to be less than this 5-$\sigma$ discovery cross section, $\sigma_{\rm Discovery}$,
\bea
\sigma_{\gamma XX} {\cal L}({\rm TeV}) &\leq & 5 \sqrt{\sigma_{\rm SM~Bkgrnd} {\cal L}{\rm(TeV)}} \cr
 &\equiv & \sigma_{\rm Discovery} {\cal L}({\rm TeV}) ,
\eea
where ${\cal L}({\rm TeV})$ is the integrated luminosity for the Tevatron-II experimental run.  This experimental constraint implies an excluded region of parameter space.  The allowed values of the brane tension will be bounded below by a function of the brane vector mass and the coupling constants' values.  As in the lepton case only co-dimension $N=1$ values will be plotted due to the mild $N^{1/8}$ dependence of the line of exclusion/accessibility.  The bounds on $F_X$ and $M_X$ are shown in Fig. 7 for different values of the coupling to the energy-momentum tensor, given by the coupling constant $\tau$, and to the extrinsic curvature related terms with coupling constants $K_1$ and $K_2$.  Fig. 8 shows excluded and allowed regions of $K_1$ and $K_2$ parameter space for fixed values of the brane tension $F_X$ and brane vector mass $M_X$ for $\tau =1$.  Table 2 summarizes the center of mass energies, the integrated luminosities and backgrounds for each collider \cite{Tevatron}.
\begin{figure}
\begin{center}
\includegraphics[height=2.2in]{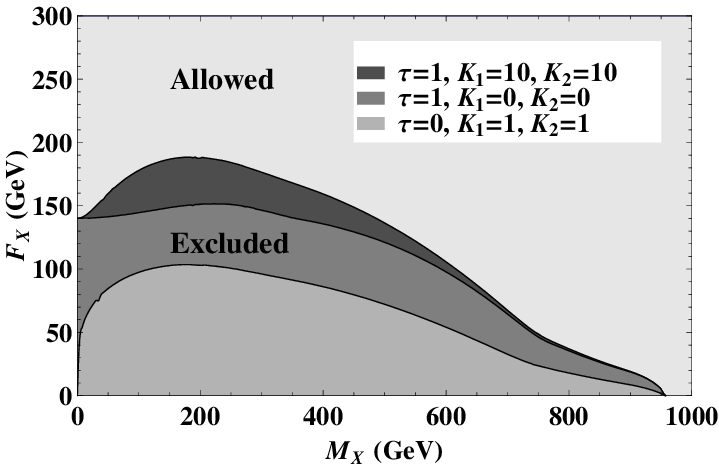}
\end{center}
\caption{Tevatron-II excluded regions of brane vector parameter space for fixed values of $\tau$, $K_1$ and $K_2$.}
\end{figure}
\begin{figure}
\begin{center}
\includegraphics[height=2.0in]{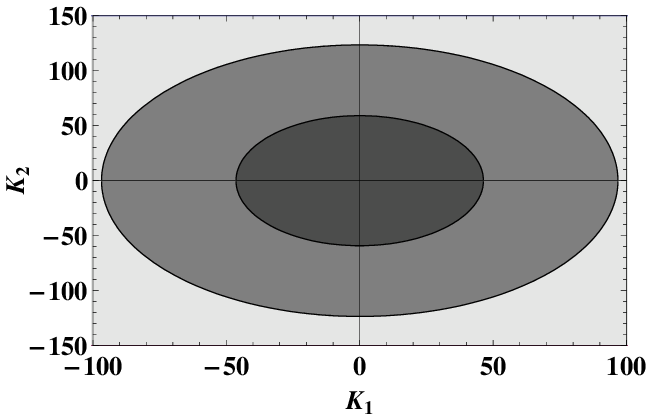}
\end{center}
\caption{Tevatron-II excluded regions of brane vector parameter space for fixed values of $F_X$, $M_X$ and $\tau =1$.  The outer lightest grey areas are excluded regions of parameter space while the 2 inner grey regions are allowed.  The innermost area corresponds to $F_X =250$ GeV and $M_X=300$ GeV.  The medium grey middle region corresponds to $F_X =300$ GeV and $M_X=300$ GeV.}
\end{figure}

As was the case with lepton colliders, for large values of the energy compared to the brane vector mass the longitudinal mode will dominate the scattering.  Once again, the equivalence theorem \cite{equivthm} dictates that the longitudinal brane vector cross section can be accurately described by that of the Nambu-Goldstone boson \cite{Kugo:1999mf}, \cite{Cembranos:2004jp}, \cite{Creminelli:2000gh}.  This effect can be seen in Fig. 9 for $\tau =1$ and $K_1 =0=K_2$ for small mass values.  The Nambu-Goldstone boson (branon) degree of freedom excluded/allowed region is shown along with the full brane vector constraints.  For small values of $M_X$ the lines of exclusion approach each other.  For larger values of the brane vector mass, the full brane vector cross section is required.  
\vspace*{0.1in}
\begin{center}
\begin{tabular}{|cccc|}
\hline
Collider&$\sqrt{s}$ (TeV)&${\cal L}~{\rm (pb^{-1})}$&$\sigma_{\rm Discovery}$ (pb)\\\hline\hline
Tevatron-II&1.96&84&0.25\\
Tevatron-6&1.96&6,000&0.029\\
LHC&14&$10^5$&0.0071\\
\hline
\end{tabular}
\end{center}
\begin{center}
{\bf Table 2.}  The colliders, their center of mass energies,\\ integrated
luminosities and discovery cross-sections.
\end{center}
\begin{figure}
\begin{center}
\includegraphics[height=2.0in]{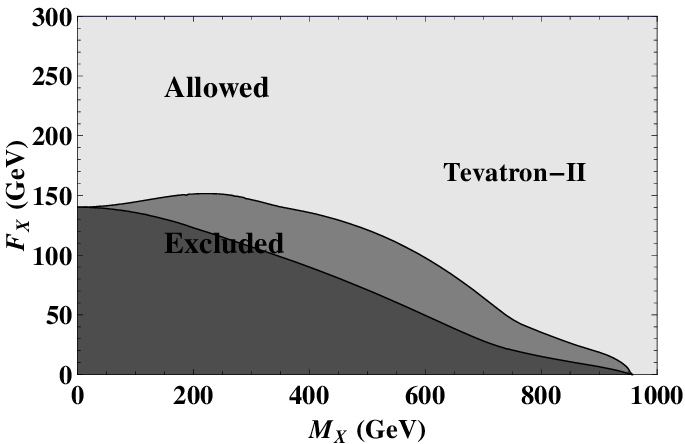}
\end{center}
\caption{Comparison of the branon and brane vector exclusion regions.  For small mass $M_X$ the equivalence theorem applies.  Branon excluded region in darker grey shading superimposed over the brane vector excluded region.}
\end{figure}

The Tevatron bounds on parameter space can be extended by its continued collection of data. It will be assumed that it can reach an integrated luminosity of $6$ ${\rm fb^{-1}}$.  Scaling the discovery cross section by the ratio of luminosities, the cross section threshold for new physics to be discovered at the Tevatron-6 is given by $\sigma_{\rm Discovery} ({\rm TeV6}) = \sigma_{\rm Discovery} ({\rm TeV}) \sqrt{\frac{{\cal L}({\rm TeV})}{{\cal L}({\rm TeV6})}}=0.029~{\rm pb}$.  The region of brane vector parameter space that will be accessible to continued Tevatron running is determined by $\sigma_{\gamma XX} \geq \sigma_{\rm Discovery} ({\rm TeV6})$.  For parameters such that the cross section for $p\bar{p} \rightarrow \gamma + XX\rightarrow \gamma + \rlap{E}{/}$ is less than the discovery cross section, $\sigma_{\gamma XX} \leq \sigma_{\rm Discovery} ({\rm TeV6})$, that region of parameter space will be inaccessible to the Tevatron-6 experiment.  Such accessible and inaccessible regions of brane vector parameter space are displayed in Figs. 10 and 11.
\begin{figure}
\begin{center}
\includegraphics[height=2.2in]{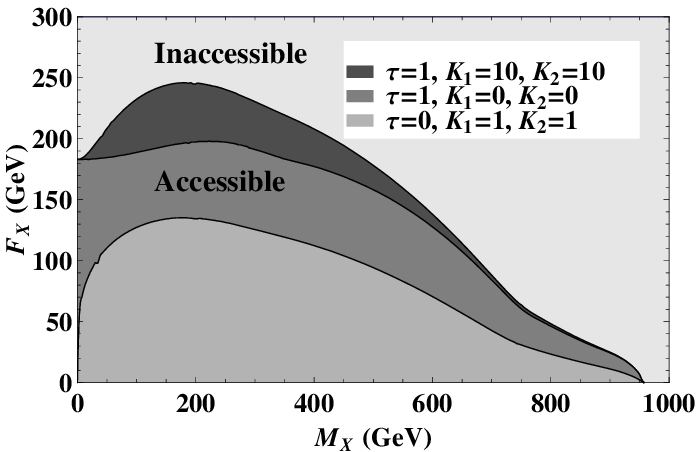}
\end{center}
\caption{Tevatron-6 accessible regions of brane vector parameter space for fixed values of $\tau$, $K_1$ and $K_2$.}
\end{figure}
\begin{figure}
\begin{center}
\includegraphics[height=2.0in]{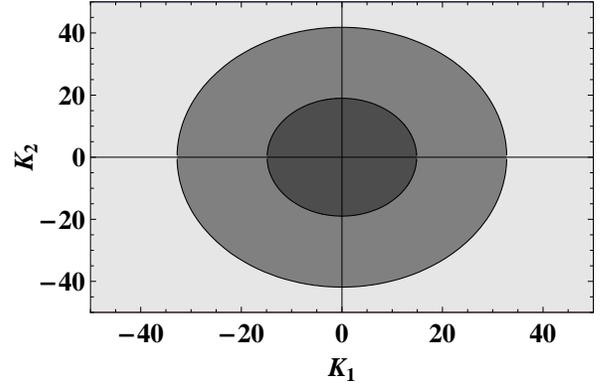}
\end{center}
\caption{Tevatron-6 accessible regions of brane vector parameter space for fixed values of $F_X$, $M_X$ and $\tau =1$.  The outer lightest grey areas are accessible regions of parameter space while the 2 inner grey regions are inaccessible.  The innermost area corresponds to $F_X =250$ GeV and $M_X=300$ GeV.  The medium grey middle region corresponds to $F_X =300$ GeV and $M_X=300$ GeV.}
\end{figure}

Much larger regions of brane vector parameter space should be accessible to the LHC \cite{Colliders}.  A conservative estimate of the discovery cross section sensitivity of the LHC is found by scaling the Tevatron background by the LHC luminosities.  Again requiring a new physics discovery to be 5-$\sigma$ above this background yields
\bea
\sigma_{\rm Discovery} ({\rm LHC}) &=& \sigma_{\rm Discovery} ({\rm TeV}) \sqrt{\frac{{\cal L}({\rm TeV})}{{\cal L}({\rm LHC})}}\cr
 &=& 0.0071~ {\rm pb}.
\eea
Hence the region of accessibility of the brane vector parameter space probed by the LHC will be for the $pp \rightarrow \gamma + XX\rightarrow \gamma + \rlap{E}{/}$ cross section to be 
\be
\sigma_{\gamma XX} \geq \sigma_{\rm Discovery} ({\rm LHC}),
\ee
and conversely, if $\sigma_{\gamma XX} \leq \sigma_{\rm Discovery} ({\rm LHC})$ the corresponding region of parameter space will be inaccessible to the LHC.  Various slices of parameter space displaying the regions of accessibility and inaccessibility are shown in Fig. 12 for fixed values of $\tau$, $K_1$, $K_2$ and $N=1$ and in Fig. 13 for fixed values of $\tau=1$, $F_X$, $M_X$ and $N=1$.
\begin{figure}
\begin{center}
\includegraphics[height=2.2in]{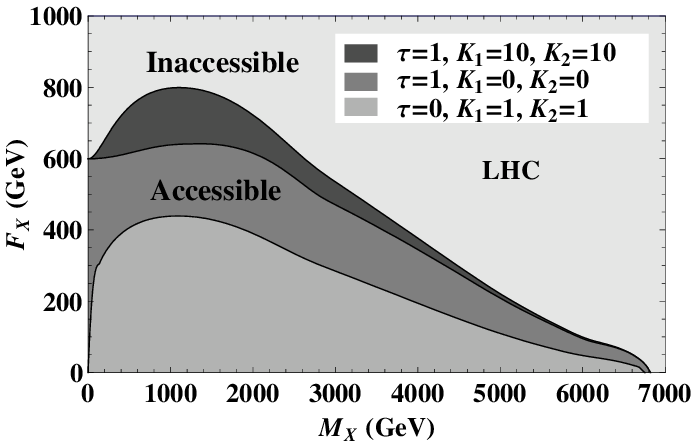}
\end{center}
\caption{LHC accessible regions of brane vector parameter space for fixed values of $\tau$, $K_1$ and $K_2$.}
\end{figure}
\begin{figure}
\begin{center}
\includegraphics[height=2.0in]{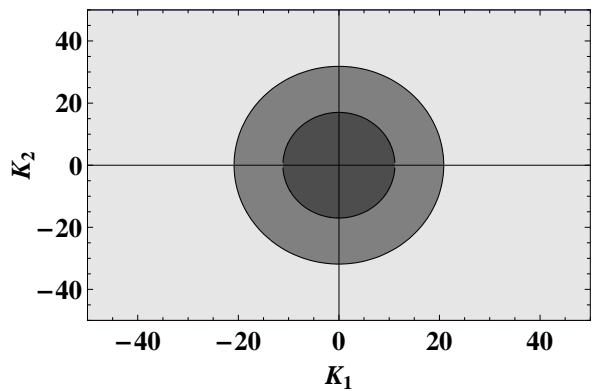}
\end{center}
\caption{LHC accessible regions of brane vector parameter space for fixed values of $\tau =1$, $F_X$ and $M_X$.  The outer lightest grey areas are accessible regions of parameter space while the 2 inner grey regions are inaccessible.  The innermost area corresponds to $F_X =700$ GeV and $M_X=2,000$ GeV.  The medium grey middle region corresponds to $F_X =800$ GeV and $M_X=2,000$ GeV.}
\end{figure}

\section{Conclusions}

The appearance of massive vector fields can signal the existence of extra space dimensions.  For a flexible brane, these brane vector fields provide the dominant coupling of brane oscillations to the Standard Model particles.  They couple pairwise to the Standard Model energy-momentum tensor directly, to the weak hypercharge field strength tensor by means of the extrinsic curvature related tensor and to the Higgs invariant bilinear.  The interaction with the Standard Model fields supplies a collider production mechanism for brane vector pairs \cite{Clark:2007wj}, \cite{Colliders}.  In particular the experimental agreement with the Standard Model for the single photon plus missing energy processes at LEP, $e^+ e^- \rightarrow \gamma + XX\rightarrow \gamma + \rlap{E}{/}$, and at the Tevatron, $p\bar{p}\rightarrow \gamma + XX \rightarrow \gamma +\rlap{E}{/}~$, was used to exclude regions of brane vector parameter space.  In addition, the same process was used to estimate the regions of brane vector parameter space accessible to the FLC, the extended running of the Tevatron and likewise the process $p{p}\rightarrow \gamma + XX \rightarrow \gamma +\rlap{E}{/}~$ for the LHC.  In the calculation of the cross section it was found that no interference occurred between the energy-momentum tensor coupling of the brane vector fields to the Standard Model and the extrinsic curvature related coupling terms to the Standard Model fields or among the extrinsic curvature coupling constants themselves.  Also the cross section depended on the effective brane tension to the eighth power.  This resulted in a mild dependence on the number of extra dimensions, $N^{1/8}$, for the line of exclusion or accessibility in the brane tension versus brane vector mass plot of parameter space and so $N=1$ was used to illustrate the regions of parameter space.  For the sections of parameter space with fixed coupling to the energy-momentum tensor, $\tau =1$, an increase in the extrinsic curvature coupling constants, $K_1$ and $K_2$, increased the brane vector production cross section and so resulted in a larger region of excluded/accessible $F_X$-$M_X$ parameter space.  In contradistinction, for larger values of the brane tension the cross section decreased, so more $K_1$-$K_2$ parameter space is allowed/inaccessible.  For low values of the brane vector mass the cross section is dominated by the production of longitudinal brane vectors as effectively described by the branon cross section in accord with the equivalence theorem.  The coupling to the hypercharge field strength tensor is through the extrinsic curvature and involves the brane vector field strength tensor as well.  Hence its contribution vanishes in this low brane vector mass limit as seen from the $F_X$-$M_X$ plots.  In general the Tevatron-II data provides tighter constraints on the parameter space as compared to LEP-II, as will the data from the Tevatron continued running to 6 ${\rm fb}^{-1}$.  The future Tevatron-6 data will provide more stringent constraints on the allowed extrinsic curvature coupling constants than the present Tevatron-II data for fixed $F_X$ and $M_X$ values as seen by comparing Fig. 8 and Fig. 11.  The data from the LHC will provide access to the largest volume of parameter space essentially doubling the reach in effective brane tension values $F_X$ and a factor of seven in brane vector mass $M_X$.  Since the brane vectors are stable they are candidates for dark matter.  It has been shown that although they elude direct detection bounds their cosmological abundance places additional bounds on the parameter space \cite{Clark:2007wj, darkmatter}.  Because of their direct coupling to the Higgs particle it is also possible that the Higgs can decay invisibly into pairs of brane vectors at rates comparable to that of the Standard Model decay modes \cite{higgsdecay}.
\newline

\noindent The work of TEC, STL and CX was supported in part by the U.S. Department of Energy under grant DE-FG02-91ER40681 (Task B).  The work of MN is supported in part by Grant-in-Aid for Scientific Research (No.~20740141) from the Ministry of Education, Culture, Sports, Science and Technology-Japan.  The work of TtV was supported in part by a Cottrell
Award from the Research Corporation and by the NSF under grant PHY-0758073.  TtV would like to thank the theoretical physics group at Purdue University for their hospitality during his sabbatical leave from Macalester College.

\newpage
\end{document}